\documentclass[12pt]{article}
 \usepackage{graphicx}
 \usepackage[cp1251]{inputenc}

\begin{document}
 \noindent {\footnotesize\it
   Astronomy Letters, 2021, Vol. 47, No 9, pp. 607--617.}

 \noindent
 \begin{tabular}{llllllllllllllllllllllllllllllllllllllllllllll}
 & & & & & & & & & & & & & & & & & & & & & & & & & & & & & & & & & & & & & &\\\hline\hline
 \end{tabular}

  \vskip 0.5cm
\centerline{\bf\large Three-Dimensional Kinematics of Classical Cepheids}
   \bigskip
   \bigskip
  \centerline
 {V. V. Bobylev \footnote [1]{e-mail: vbobylev@gaoran.ru} and A. T. Bajkova}
   \bigskip

  \centerline{\small\it Pulkovo Astronomical Observatory, Russian Academy of Sciences,}

  \centerline{\small\it Pulkovskoe sh. 65, St. Petersburg, 196140 Russia}
 \bigskip
 \bigskip
 \bigskip

{\bf Abstract}---A linear Ogorodnikov–Milne model is applied to study the three-dimensional kinematics of classical Cepheids in the Milky Way. A sample of 832 classical Cepheids from Mr\'oz et al. (2019) with distances, line-of-sight velocities, and proper motions from the Gaia DR2 catalogue is used. The Cepheid space velocities have been freed from the differential Galactic rotation found by us previously based on a nonlinear rotation model. Based on a complete Ogorodnikov–Milne model, involving the line-of-sight velocities and proper motions of stars, we have estimated the angular velocity of rotation around the
Galactic $y$ axis, $\Omega_y=0.64\pm0.17$~km s$^{-1}$ kpc$^{-1}$. We think that this rotation is associated with the warp of the Galactic thin disk. Our calculations using only the proper motions of Cepheids under the assumption of no deformations due to the disk warp have shown the presence of a residual rotation around the $y$ axis with an angular velocity $\Omega_y=0.54\pm0.15$~km s$^{-1}$ kpc$^{-1}$ and the presence of a positive rotation around the $x$ axis with an angular velocity $\Omega_x=0.33\pm0.10$~km s$^{-1}$ kpc$^{-1}$.


 \subsection*{INTRODUCTION}
A large-scale warp of the thin disk is observed in the Galaxy. This effect was first revealed by HI observations (Westerhout 1957). A rise of hydrogen above the Galactic plane is observed in the first and second Galactic quadrants, while in the third and
fourth ones, on the contrary, it is located below the
Galactic plane, with the warp amplitude increasing
toward the Galactic anticenter. At present, the disk
warp is confirmed by various data---from the distribution
of neutral (Kalberla and Dedes 2008) and ionized
(Russeil 2003; Cersosimo et al. 2009) hydrogen, interstellar
dust (Drimmel and Spergel 2001), pulsars
(Yusifov 2004), OB stars (Miyamoto and Zhu 1998;
Drimmel et al. 2000), red giant clump stars (L\'opez-Corredoira et al. 2002b; Momany et al. 2006), various stars (Cheng et al. 2020), and Cepheids (Fernie
1968; Berdnikov 1987; Bobylev 2013a; Skowron et al. 2019b).

For example, according to Bailin (2003), the warp is induced by the gravitational torque that is exerted on the Galactic disk by the nearest dwarf galaxies, in particular, the Large Magellanic Cloud (Bekki 2012). Other hypotheses proposed to explain the origin of the warp are also known: the interaction between the disk and a nonspherical dark matter
halo (Sparke and Casertano 1988); the interaction
of the disk with a circumgalactic flow produced by
high-velocity hydrogen clouds that resulted from
mass transfer between theGalaxy and theMagellanic
Clouds (Tsuchiya 2002; Olano 2004); an intergalactic flow (L\'opez-Corredoira et al. 2002a); and the interaction with an intergalactic magnetic field (Battaner et al. 1990).

Various methods of analyzing the vertical velocities of stars were applied in searching for a connection between the kinematics of stars and the warp. For example, Drimmel et al. (2000) used the proper motion components $\mu_b$ of OB stars from the Hipparcos (1997) catalogue for this purpose. In particular,
these authors showed that there is a systematic component
$\sim$10--15~km s$^{-1}$ in the vertical velocities of
OB stars at Galactocentric distances $R\sim11$ kpc. A fairly high precession rate of the warp, from $-13$ to $-25$~km s$^{-1}$ kpc$^{-1}$, depending on the adopted model of errors in the data, was detected.

Skowron et al. (2019b) constructed the vertical velocity distribution of Cepheids in the Milky Way disk. They found large-scale vertical motions with amplitudes of 10--20~km s$^{-1}$, such that Cepheids located in the northern warp (approximately in the first
and second Galactic quadrants) exhibit a large positive vertical velocity (toward the north Galactic pole), whereas those in the southern warp (approximately in the third and fourth quadrants) exhibit a negative vertical velocity (toward the south Galactic pole).

It is interesting to note the results obtained with a more complex Ogorodnikov–Milne model, where the peculiar solar motion parameters, the Galactic rotation parameters (rotation around the Galactic $z$ axis), the deformation parameters in the $xz$ and $yz$ planes,
and the angular velocities of rotation around the $z$ and $y$ axes act as the unknowns to be determined. Using this method, Miyamoto and Zhu (1998) found rotation of a system of O-B5 stars (from their proper motions) around the $x$ axis with an angular velocity of
$+3.8\pm1.1$~km s$^{-1}$ kpc$^{-1}$, while Bobylev (2010) found
rotation around the x axis with an angular velocity of
about $-4$~km s$^{-1}$ kpc$^{-1}$ from the proper motions of
red giant clump stars. We see that there is no good
agreement in the results of the analysis. Moreover, having studied $\sim$200 long-period Cepheids by this method, Bobylev (2013b) found a fairly large rotation around the $x$ axis with an angular velocity of about $-15$~km s$^{-1}$ kpc$^{-1}$.

L\'opez-Corredoira et al. (2014) studied the vertical velocities of Galactic disk stars in the $R$ range 5--16 kpc. For this purpose, they used the proper motions of stars from the PPMXL catalogue (R\"oser et al. 2010). Their main goal was to ascertain whether the warp is a long-lived or a transient feature. As a result, these authors concluded that the lifetime of the warp is $\sim$100 Myr.

Based on data from the Gaia DR2 (Brown et al. 2018) and 2MASS (Skrutskie et al. 2006) catalogues, Poggio et al. (2018) mapped the kinematic signature of the Galactic stellar warp out to a distance of 7 kpc from the Sun. For this purpose, they analyzed the space velocities of $\sim$600 thousand young main-sequence stars and $\sim$13 million giants. The large-scale kinematics of all these stars was shown to have a clear signature of the warp, apparent as a gradient of 5--6~km s$^{-1}$ in their vertical velocities,
$\partial W/\partial x$ and $\partial W/\partial y$, from 8 to 14 kpc in Galactic radius.
It is also interesting to note the paper by Poggio et al. (2020), where the precession rate of the warp was estimated by analyzing $\sim$12 million giants from the Gaia DR2 catalogue to be $10.86\pm0.03 (stat.)\pm3.20 (syst.)$ km s$^{-1}$ kpc$^{-1}$ in the direction of Galactic rotation.

The goal of this paper is to study the three-dimensional kinematics of classical Cepheids in the Milky Way using the Ogorodnikov–Milne model. In particular, it is interesting to ascertain the pattern of motions in the $xz$ and $yz$ planes and to determine the angular velocities of rotation around the $x$ and y axes. For this purpose, we used a sample from
Mr\'oz et al. (2019), where the distances, line-of-sight velocities, and proper motions from the Gaia DR2 catalogue are given for 832 classical Cepheids.

 \section*{DATA}
In this paper we use data on classical Cepheids from Skowron et al. (2019a) and Mr\'oz et al. (2019). These Cepheids were observed within the fourth stage of the OGLE (Optical Gravitational Lensing Experiment, Udalski et al. 2015) program. The catalogue of Skowron et al. (2019a) contains distance, age, pulsation period estimates and photometric data for 2431 Cepheids. Their apparent magnitudes lie in the range $11^m<I<18^m$. Therefore, a deficit of bright and well-studied Cepheids known from earlier observations is observed here.

The heliocentric distances to 2214 Cepheids, $r$,
were calculated by Skowron et al. (2019a) based on
the period--luminosity relation. The specific relation
used by them was refined by Wang et al. (2018) from
the light curves of Cepheids in themid-infrared range,
where the interstellar extinction is much lower than
that in the optical one. The Cepheid ages in Skowron
et al. (2019a) were estimated by the technique developed by Anderson et al. (2016), where the stellar rotation periods and metallicity indices were taken into account.

The catalogue of Mr\'oz et al. (2019) contains 832 classical Cepheids from from the list of Skowron et al. (2019a). The proper motions in it were
copied from the Gaia DR2 catalogue; the line-ofsight
velocities are given for all 832 stars. Bobylev
et al. (2021) provided the Cepheids from the catalogue of Mr\'oz et al. (2019) with the age estimates from the catalogue of Skowron et al. (2019a).

From kinematic data on Cepheids Mr\'oz et al. (2019) constructed the Galactic rotation curve in the range of distances $R:4-20$ kpc. Based on a large sample of Cepheids, Ablimit et al. (2020) refined the parameters of the Galactic gravitational potential
and obtained a new estimate of the Galactic mass.
Bobylev et al. (2021) showed that even old Cepheids
retain kinematic memory of their birthplace. The
sample of 832 Cepheids with kinematic data from
Mr\'oz et al. (2019) is our working sample.

 \section*{METHOD}
 \subsection*{Linear Ogorodnikov–Milne Model}
The following quantities are known from observations: the right ascension and declination, $\alpha$ and $\delta$; the proper motions in right ascension and declination,
$\mu_\alpha\cos \delta$ and $\mu_\delta$; and the line-of-sight velocity $V_r.$
We pass from $\alpha$ and $\delta$ to the Galactic longitude and latitude, $l$ and $b$; we calculated the heliocentric distances $r$ for the Cepheids based on the period--luminosity relation; we convert the observed proper motions to the proper motions in the Galactic coordinate system, $\mu_l\cos b$ and $\mu_b$. As a result, we have three components of the stellar space velocity: $V_r$ and two projections of the tangential velocity,
$V_l=k\,r\,\mu_l\cos b$ and $V_b=k\,r\,\mu_b$, where $k=4.74$~km s$^{-1}$, $V_r, V_l,$ and $V_b$ are expressed in km s$^{-1}$ (the proper motions and heliocentric distances are given in mas yr$^{-1}$ (milliarcseconds per year) and kpc, respectively).

We use a rectangular Galactic coordinate system with the axes directed away from the observer toward the Galactic center (the $x$ axis or axis 1), in the direction
of Galactic rotation (the $y$ axis or axis 2), and toward the north Galactic pole (the $z$ axis or axis 3).

In the linear Ogorodnikov–Milne model (Ogorodnikov 1965) the observed velocity ${\bf V}(r)$ of a star with a heliocentric radius vector ${\bf r}$ is described, to terms of the first order of smallness $r/R_0\ll 1$, by the vector equation
\begin{equation}
 {\bf V}(r)={\bf V}_\odot+M{\bf r}+{\bf V'},
 \label{eq-1}
 \end{equation}
where ${\bf V}_\odot(U_\odot,V_\odot,W_\odot)$ is the peculiar solar velocity relative to the group of stars under consideration, $M$ is the displacement matrix (tensor) whose
components are the partial derivatives of the velocity
${\bf u}(u_1,u_2,u_3)$ with respect to the distance ${\bf r}(r_1, r_2, r_3)$,
where ${\bf u}={\bf V}(R)-{\bf V}(R_0)$, while $R$ and $R_0$ are the
heliocentric distances of the star and the Sun, respectively.
Then,
\begin{equation}
 M_{pq}={\left(\frac{\partial u_p} {\partial r_q}\right)}_\circ, \quad p,q=1,2,3,
 \label{eq-2}
 \end{equation}
where zero means that the derivatives are taken at $R=R_0$, while the subscripts $p$ and $q$ denote the coordinate axis numbers, $\bf V'$ is the residual stellar velocity, by which we mean the stellar velocity after the subtraction of the peculiar solar motion (${\bf V}_\odot$) and the linear dependences described by the displacement matrix $M.$ Note that here we adhere to the notation introduced by Clube (1972).

All nine elements of the matrix $M$ can be determined using three components of the observed
velocities—the line-of-sight velocity $V_r$ and the velocities along the Galactic longitude $V_l$ and the Galactic latitude $V_b$:
 \begin{equation}
  \begin{array}{lll}
 V_r=-U_{\odot}\cos b\cos l
   -V_{\odot}\cos b\sin l-W_{\odot}\sin b\\
   +r[\cos^2 b\cos^2 l M_{11}+\cos^2 b\cos l\sin l M_{12}+\cos b\sin b \cos l  M_{13}\\
   +\cos^2 b\sin l\cos l M_{21}+\cos^2 b\sin^2 l M_{22}+\cos b\sin b\sin l M_{23}\\
   +\sin b\cos b\cos lM_{31}+\cos b\sin b\sin lM_{32} +\sin^2 b M_{33}], \label{eq-3}
 \end{array}
 \end{equation}
 \begin{equation}
 \begin{array}{lll}
  V_l= U_\odot\sin l-V_\odot\cos l\\
 +r [-\cos b\cos l\sin l  M_{11} -\cos b\sin^2 l M_{12}-\sin b \sin l  M_{13}\\
 +\cos b\cos^2 l M_{21}+\cos b\sin l\cos l M_{22}+\sin b\cos l  M_{23} ], \label{eq-4}
\end{array}
 \end{equation}
 \begin{equation}
 \begin{array}{lll}
 V_b=U_\odot\cos l\sin b+V_\odot\sin l\sin b-W_\odot\cos b\\
 +r [-\sin b\cos b\cos^2 l M_{11}-\sin b\cos b\sin l \cos l M_{12}-\sin^2 b \cos l M_{13}\\ -\sin b\cos b\sin l\cos l M_{21}-\sin b\cos b\sin^2 l M_{22}-\sin^2 b\sin l M_{23}\\
 +\cos^2 b\cos l M_{31} +\cos^2 b\sin l M_{32}+\sin b\cos b  M_{33} ].
   \label{eq-5}
  \end{array}
 \end{equation}
To estimate the velocities $(U,V,W)_\odot$ and the elements of the matrix $M$, the system of conditional equations (3)--(5) is solved by the least-squares method (LSM). The solution is sought with weights of the form
\begin{equation}\label{weights}
 w_r =S_0/\sqrt{S_0^2+\sigma_{V_r}^2}, \quad
 w_l =S_0/\sqrt{S_0^2+\sigma_{V_l}^2}, \quad
 w_b =S_0/\sqrt{S_0^2+\sigma_{V_b}^2},
\end{equation}
where $\sigma_{V_r}, \sigma_{V_l},$ and $\sigma_{V_b}$ are the errors in the corresponding
observed velocities, $S_0$ is the ``cosmic'' dispersion. $S_0$ is comparable to the root-mean-square residual $\sigma_0$ (the error per unit weight) calculated by solving the conditional equations (3)--(5) and taken to be 12 km s$^{-1}$ in this paper.

The matrix $M$ is divided into symmetric,
 $M^{\scriptscriptstyle+}$ (local deformation tensor) and antisymmetric,
 $M^{\scriptscriptstyle-}$ (rotation tensor), parts:
 \begin{equation}
 \renewcommand{\arraystretch}{2.2}
  \begin{array}{lll}\displaystyle
 M_{\scriptstyle pq}^{\scriptscriptstyle+}=
 {1\over 2}\left( \frac{\partial u_{p}}{\partial r_{q}}+
 \frac{\partial u_{q}}{\partial r_{p}}\right)_\circ,\quad
 \displaystyle
 M_{\scriptstyle pq}^{\scriptscriptstyle-}=
 {1\over 2}\left(\frac{\partial u_{p}}{\partial r_{q}}-
 \frac{\partial u_{q}}{\partial r_{p}}\right)_\circ, \quad
  p,q=1,2,3,
 \label{eq-456}
 \end{array}
 \end{equation}
where zero means that the derivatives are taken at $R=R_0$. $M_{\scriptscriptstyle32}^{\scriptscriptstyle-},
  M_{\scriptscriptstyle13}^{\scriptscriptstyle-},
  M_{\scriptscriptstyle21}^{\scriptscriptstyle-}$ are the components of
the solid-body rotation vector of a small solar neighborhood
around the $x, y,$ and $z$ axes, respectively. In accordance with our chosen rectangular coordinate system, the positive rotations are those from axis 1 to 2 ($\Omega_z$), from axis 2 to 3 ($\Omega_x$), and from axis 3 to 1 ($\Omega_y$):
 \begin{equation}
 M^{\scriptscriptstyle-}= \pmatrix
  {         0&-\Omega_z &~~\Omega_y\cr
   ~~\Omega_z&         0&-\Omega_x\cr
    -\Omega_y&~~\Omega_x&         0\cr}.
 \label{Omega-0}
 \end{equation}
The components of the rotation tensor are calculated
from the elements of the matrix $M$ as follows:
\begin{equation}\label{B-xyz}
    M_{\scriptscriptstyle32}^{\scriptscriptstyle-}=
0.5(M_{\scriptscriptstyle32}-M_{\scriptscriptstyle23}),\quad
    M_{\scriptscriptstyle13}^{\scriptscriptstyle-}=
0.5(M_{\scriptscriptstyle13}-M_{\scriptscriptstyle31}),\quad
    M_{\scriptscriptstyle21}^{\scriptscriptstyle-}=
0.5(M_{\scriptscriptstyle12}-M_{\scriptscriptstyle21}).
\end{equation}
Each of the quantities
 $M_{\scriptscriptstyle12}^{\scriptscriptstyle+},
 M_{\scriptscriptstyle13}^{\scriptscriptstyle+},
 M_{\scriptscriptstyle23}^{\scriptscriptstyle+}$ describes
the deformation in the corresponding plane. They are calculated from the elements of the matrix $M$ as follows:
\begin{equation}\label{A-xyz}
    M_{\scriptscriptstyle12}^{\scriptscriptstyle+}=
0.5(M_{\scriptscriptstyle12}+M_{\scriptscriptstyle21}),\quad
    M_{\scriptscriptstyle13}^{\scriptscriptstyle+}=
0.5(M_{\scriptscriptstyle13}+M_{\scriptscriptstyle31}),\quad
    M_{\scriptscriptstyle23}^{\scriptscriptstyle+}=
0.5(M_{\scriptscriptstyle23}+M_{\scriptscriptstyle32}).
\end{equation}
The diagonal components of the local deformation
tensor
  $M_{\scriptscriptstyle11}^{\scriptscriptstyle+},
  M_{\scriptscriptstyle22}^{\scriptscriptstyle+},
  M_{\scriptscriptstyle33}^{\scriptscriptstyle+},$ coincide with the corresponding
diagonal elements of the matrix $M.$ They
describe the general local contraction or expansion of
the entire stellar system (divergence). In particular,
it is interesting to estimate the three-dimensional
expansion/contraction effect:
 \begin{equation}
 K_{xyz}= (M_{\scriptscriptstyle11}+
           M_{\scriptscriptstyle22}+
           M_{\scriptscriptstyle33})/3.
 \label{K-XYZ}
 \end{equation}
The rectangular components of the stellar space velocities
are calculated from the formulas
\begin{equation}\label{UVW}
 \begin{array}{lll}
U=V_r \cos l \cos b-V_l \sin l-V_b \cos l \sin b,\\
V=V_r \sin l \cos b+V_l \cos l-V_b \sin l \sin b,\\
W=V_r \sin b+V_b \cos b.
\end{array}
\end{equation}

\subsection*{Formation of the Residual Velocities}
The parameters of the Galactic rotation curve were found in Bobylev et al. (2021) by expanding the angular velocity of Galactic rotation $\Omega$ into a series to
terms of the $i$th order of smallness in $r/R_0,$ where $\Omega^i_0$ are the corresponding derivatives of the angular velocity.

The residual velocities of the Cepheids are calculated by taking into account the peculiar solar velocity $U_\odot, V_\odot,$ $W_\odot$ and the differential Galactic rotation in
the following form:
\begin{equation}
 \begin{array}{lll}
 V_r=V^*_r
 -[-U_\odot\cos b\cos l-V_\odot\cos b\sin l-W_\odot\sin b\\
 +R_0(R-R_0)\sin l\cos b\Omega^\prime_0
 +0.5R_0(R-R_0)^2\sin l\cos b\Omega^{\prime\prime}_0+\ldots],
 \label{EQU-1}
 \end{array}
 \end{equation}
 \begin{equation}
 \begin{array}{lll}
 V_l=V^*_l -[U_\odot\sin l-V_\odot\cos l-r\Omega_0\cos b\\
 +(R-R_0)(R_0\cos l-r\cos b)\Omega^\prime_0
 +0.5(R-R_0)^2(R_0\cos l-r\cos b)\Omega^{\prime\prime}_0+\ldots],
 \label{EQU-2}
 \end{array}
 \end{equation}
  \begin{equation}
 \begin{array}{lll}
 V_b=V^*_b -[U_\odot\cos l\sin b + V_\odot\sin l\sin b-W_\odot\cos b\\
 -R_0(R-R_0)\sin l\sin b\Omega^\prime_0
 -0.5R_0(R-R_0)^2\sin l\sin b\Omega^{\prime\prime}_0 -\ldots],
 \label{EQU-3}
 \end{array}
 \end{equation}
where the velocities $V^*_r,V^*_l,V^*_b,$ on the righthand
sides of the equations are the initial ones, while
the corrected velocities $V_r, V_l,$ and $V_b$ using which
the residual velocities $U, V,$ and $W$ can be calculated
from Eqs. (12), are on the left-hand sides of the
equations.

The expressions that were used in searching for the Galactic rotation parameters are in the square brackets on the right-hand sides of Eqs. (13)--(15) with their signs. In particular, based on Cepheids from Mr\'oz et al. (2019), Bobylev et al. (2021) found $(U_\odot,V_\odot,W_\odot)=(8.53,14.88,6.09)\pm(0.47,0.63,0.45)$~km s$^{-1}$ and
 \begin{equation}
 \label{solut-99}
 \begin{array}{lll}
        \Omega_0=-28.71\pm0.15~\hbox{km s$^{-1}$ kpc$^{-1}$},\\
    \Omega^{'}_0=~~3.957\pm0.044~\hbox{km s$^{-1}$ kpc$^{-2}$},\\
   \Omega^{''}_0=-0.871\pm0.033~\hbox{km s$^{-1}$ kpc$^{-3}$},\\
  \Omega^{'''}_0=~~0.153\pm0.013~\hbox{km s$^{-1}$ kpc$^{-4}$},\\
   \Omega^{IV}_0=-0.013\pm0.002~\hbox{km s$^{-1}$ kpc$^{-5}$}
 \end{array}
 \end{equation}
for the adopted $R_0=8.0$~kpc. With these parameters Bobylev et al. (2021) constructed Fig. 4c. In this paper the residual velocities of the Cepheids are formed with the parameters of the angular velocity of Galactic rotation (16).

 \begin{figure} [t] {\begin{center}
  \includegraphics[width=150mm]{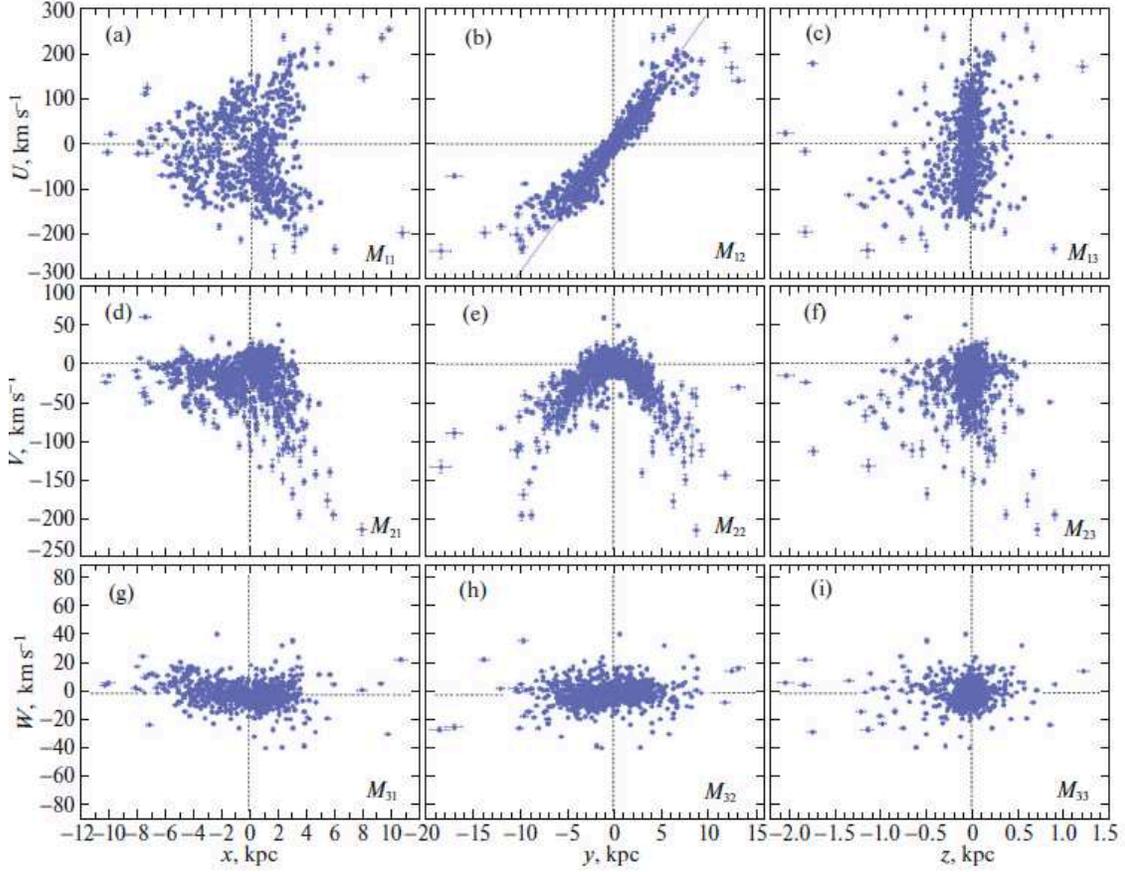}
 \caption{
Velocities $U,V,W$ versus heliocentric rectangular coordinates $x,y,z$. The corresponding designations of the deformation matrix $M_{p,q}$, are given on each panel; on panel (b) the gray line indicates the dependence $U=M_{12}\cdot y$, where $M_{12}=28.71$~km s$^{-1}$ kpc$^{-1}$. The value of the parameter $M_{12}$ is in the confidence interval~[26.74,30.687]~km s$^{-1}$ kpc$^{-1}$ with a 95\% probability.
 }
 \label{f-all-1}
 \end{center} }
 \end{figure}

 \subsection*{Constraints}
To get rid of several outlying residual Cepheid velocities, we use the following constraints:
 \begin{equation}
 \begin{array}{rcl}
    |U|<80~\hbox{\rm km s$^{-1}$},\quad
    |V|<80~\hbox{\rm km s$^{-1}$},\quad
    |W|<60~\hbox{\rm km s$^{-1}$},
  \label{cut}
 \end{array}
 \end{equation}
where the velocities $U, V,$ and $W$ are the residual
ones, i.e., corrected for the Galactic rotation.

In Fig. 1 the velocities $U, V,$ and $W$ are plotted against the heliocentric rectangular coordinates $x, y,$ and $z.$ The velocities $U, V,$ and $W$ were calculated using
the constraints (17), based on which we discarded only 16 stars from the general list.

Figure 1b shows the dependence $U=M_{12}\cdot y$, where $M_{12}$ corresponds to the angular velocity of Galactic rotation around the $z$ axis (see (8)), whose value we use in (16),
$M_{12}=\partial U/\partial y=-\Omega_z=28.71$~km s$^{-1}$ kpc$^{-1}$. The points on the plot are seen to closely follow this linear dependence in a wide range, $-10$~kpc~$<y<10$~kpc. However, in other cases, especially in Fig.~1e, the velocity distribution is distinctly nonlinear in pattern.

We tested the statistical significance of the regression in Fig.~1b using Student's $t$-test. The dependence turned out to be statistically significant. The confidence interval in which the regression parameter
found is located with a 95\% probability is given in the
caption to the figure.

As can be seen from Fig. 1, the velocities $U$ and $V$ vary over a very wide range. Therefore, to study the gradients
$\partial U/\partial z$ and $\partial V/\partial z$, it is necessary to
carefully take into account the differential Galactic
rotation. It can also be seen that the velocities $W$ have a considerably smaller amplitude and do not correlate with the Galactic rotation.

{\begin{table}[t] \caption[]{\small\baselineskip=1.0ex\protect
 Kinematic parameters of the Ogorodnikov–Milne model
 }
\begin{center}
\label{t1}
\begin{tabular}{|c|r|r|r|}\hline
  Parameter & $R<12$~kpc & $R<14$~kpc & All Cepheids \\\hline

 $N_\star$        &  685 &  778 &  816 \\
 $\sigma_0$,~km s$^{-1}$ & 12.44 & 12.46 & 12.47 \\

 $U_\odot$,~km s$^{-1}$ & $ 8.28\pm0.50$ & $ 8.53\pm0.46$ & $ 8.61\pm0.45$ \\
 $V_\odot$,~km s$^{-1}$ & $15.17\pm0.50$ & $14.68\pm0.46$ & $14.57\pm0.46$ \\
 $W_\odot$,~km s$^{-1}$ & $ 6.01\pm0.49$ & $ 5.56\pm0.45$ & $ 5.47\pm0.45$ \\
           &&&  \\
 $M_{11}$,~km s$^{-1}$ kpc$^{-1}$ & $ 0.66\pm0.26$ & $ 0.74\pm0.19$ & $ 0.68\pm0.17$ \\
 $M_{12}$,~km s$^{-1}$ kpc$^{-1}$ & $-0.13\pm0.15$ & $-0.02\pm0.14$ & $ 0.00\pm0.13$ \\
 $M_{13}$,~km s$^{-1}$ kpc$^{-1}$ & $-4.40\pm3.10$ & $-5.02\pm2.33$ & $-8.11\pm1.53$ \\

 $M_{21}$,~km s$^{-1}$ kpc$^{-1}$ & $-0.18\pm0.26$ & $-0.27\pm0.19$ & $-0.27\pm0.17$ \\
 $M_{22}$,~km s$^{-1}$ kpc$^{-1}$ & $-0.38\pm0.15$ & $-0.18\pm0.14$ & $-0.18\pm0.13$ \\
 $M_{23}$,~km s$^{-1}$ kpc$^{-1}$ & $ 5.24\pm3.10$ & $ 7.54\pm2.28$ & $ 7.36\pm1.52$ \\

 $M_{31}$,~km s$^{-1}$ kpc$^{-1}$ & $-0.24\pm0.25$  & $-0.58\pm0.19$ & $-0.64\pm0.17$ \\
 $M_{32}$,~km s$^{-1}$ kpc$^{-1}$ & $ 0.24\pm0.14$  & $ 0.19\pm0.13$ & $ 0.25\pm0.13$ \\
 $M_{33}$,~km s$^{-1}$ kpc$^{-1}$ & $ 6.61\pm2.97$  & $ 2.42\pm2.18$ & $ 2.92\pm1.46$ \\
            &&&  \\
 $K_{xyz}$,~km s$^{-1}$ kpc$^{-1}$ & $2.31\pm1.00$ & $ 1.00\pm0.74$ & $ 1.13\pm0.50$ \\\hline
\end{tabular}
\end{center}
\end{table}
}
 \begin{figure} [t] {\begin{center}
  \includegraphics[width=145mm]{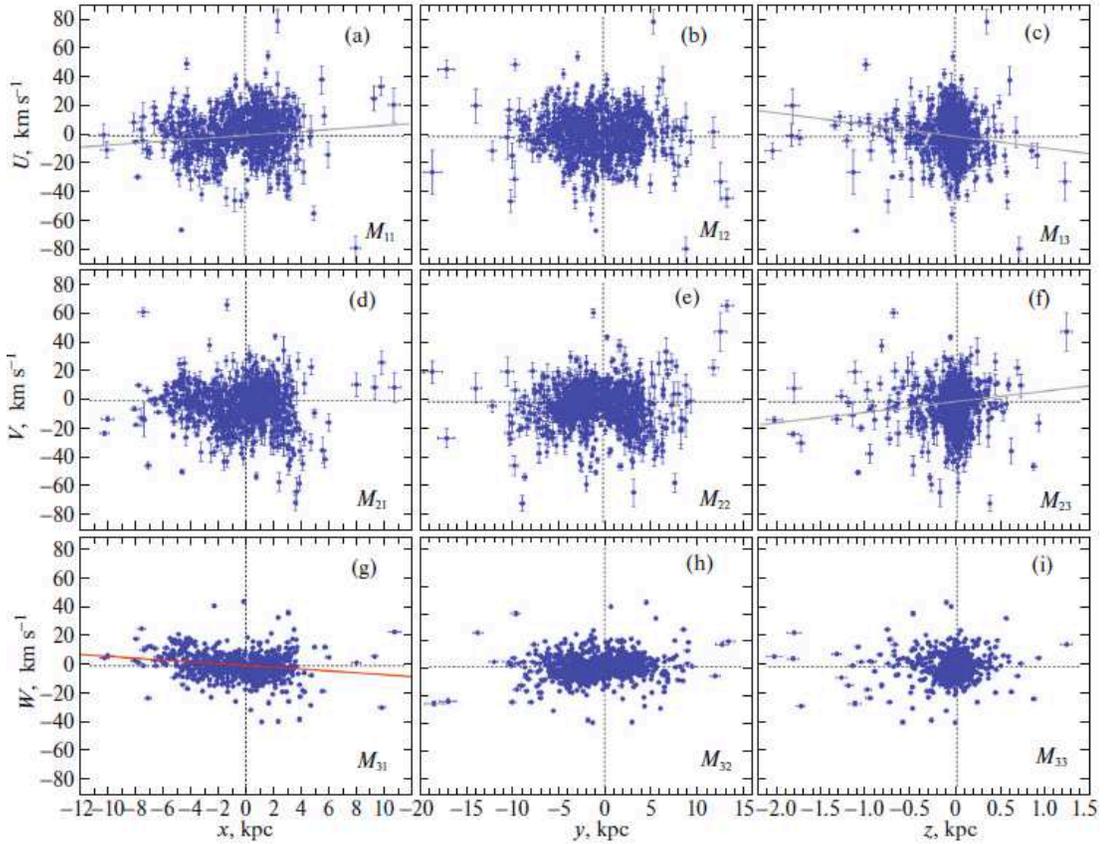}
 \caption{
Velocities $U,V,W$ corrected for the Galactic rotation versus heliocentric rectangular coordinates $x,y,z$. The corresponding designations of the deformation matrix $M_{p,q}$ are given on each panel; panel (a) shows the dependence $U=M_{11}\cdot x$, where $M_{11}=0.68$ km s$^{-1}$ kpc$^{-1}$ with the confidence interval [63, 73] km s$^{-1}$ kpc$^{-1}$; panel (c) presents the dependence $U=M_{13}\cdot z$, where $M_{13}=-8.11$ km s$^{-1}$ kpc$^{-1}$ with the confidence interval [-8.68, -7.54] km s$^{-1}$ kpc$^{-1}$; panel (f) presents the dependence $V=M_{23}\cdot z$ where $M_{23}=7.36$ km s$^{-1}$ kpc$^{-1}$  with the confidence interval [6.73, 7.99] km s$^{-1}$ kpc$^{-1}$; panel (g) presents the dependenceW $W=M_{31}\cdot x$ where $M_{31}=-0.64$ km s$^{-1}$ kpc$^{-1}$ with the confidence interval [-0.69, -0.59] km s$^{-1}$ kpc$^{-1}$. }
 \label{f-all-2}
 \end{center} }
 \end{figure}

\section*{RESULTS AND DISCUSSION}
\subsection*{Complete Model}
Table 1 gives the parameters of the linear Ogorodnikov-Milne model found through the LSM solution of the system of conditional equations (3)--(5). In this
case, the velocities $V_r,V_l,$ and $V_b$ were corrected for
the differential Galactic rotation using Eqs. (13)--(15)
and the parameters (16).

The number of stars used $N_\star$, the error per unit weight $\sigma_0$ found through the LSM solution of the system of conditional equations (3)--(5), the components
of the peculiar solar velocity $(U,V,W)_\odot$, and all nine
elements of the matrix $M$ (2) are given in the upper part of the table.

We expect that the kinematic effects associated with the warpmustmanifest themselves at significant heliocentric distances, approximately at $r>6$~kpc ($R>14$~kpc). The maps of the distribution of objects far from the Sun (Russeil 2003; Momany et al. 2006;
Kalberla and Dedes 2008; Skowron et al. 2019b), the
vertical velocity distribution of distant stars (Poggio et al. 2018), and warp models (L\'opez-Corredoira et al. 2002b; Yusifov 2004; Chrob\'akov\'a and L\'opez-Corredoira 2021) underlie these expectations. It is particularly interesting to note the paper by Poggio
et al. (2018), where it is clearly seen that the vertical velocities of distant stars begin to increase dramatically
at $R>14$~kpc. Therefore, the calculations
were performed for three cases of the Galactocentric
distance: $R<12$~kpc, $R<14$~kpc, and at any $R$.

The corresponding results are given in three
columns. Thus, we have three samples of Cepheids
at different distances from the Galactic center.

The velocities $(U,V,W)_\odot$ are the group velocity
(with the opposite sign) of the sample of Cepheids
under consideration. This includes the peculiar solar
motion relative to the local standard of rest, the
perturbations from the spiral density wave, and the
influence of the so-called asymmetric drift (lagging
behind the circular rotation velocity with sample age)
on the velocity $V_\odot$

At present, the components of the peculiar solar velocity relative to the local standard of rest are believed to have been best determined by Sch\"onrich
et al. (2010): $(U,V,W)_\odot=(11.1,12.2,7.3)\pm(0.7,0.5,0.4)$~km s$^{-1}$. We see that the group velocities found by us are in good agreement with this estimate. The slight difference of about 2.5~km s$^{-1}$ in the velocities $V_\odot$ can be explained by the influence of the asymmetric drift. Note that Bobylev et al. (2021)
found close values, $(U,V,W)_\odot=(10.1,13.6,7.0)\pm(0.5,0.6,0.4)$~km s$^{-1}$, while analyzing the same sample of Cepheids based on a nonlinear Galactic rotation model.

As can be seen from the first column ($R<12$~kpc) of the table, there is no element of the matrix $M$ that differs significantly from zero. In the second column ($R<14$~kpc) of the table there are two elements that differ significantly from zero, $M_{11}$ and $M_{23}$. In the third column (at any $R$) of the table there are four such elements:
$M_{11},$ $M_{13},$ $M_{23}$ and $M_{31}$.

In Fig. 2 the velocities $U, V,$ and $W$ corrected for the Galactic rotation are plotted against the coordinates $x, y,$ and $z.$ The sample of Cepheids corresponds to the case where no constraint was imposed on the distance $R.$ The four dependences for
$M_{11},$ $M_{13},$ $M_{23}$ and $M_{31}$ corresponding to the gradients
found, whose values are given in the last column of the table, are shown.

The four nonzero estimates of the regression coefficients were tested for statistical significance using Student's $t$-test. They all turned out to be statistically
significant. The corresponding confidence intervals in which the regression parameters are located with a 95\% probability are given in the caption to Fig. 2.

{\bf The $XY$ plane.} Consider the displacement tensor describing the residual rotation around the $z$ axis. Let us designate this tensor as $M_{xy},$ because its elements are the partial derivatives of the velocities $U$ and $V$ with respect to $x$ and $y:$
 \begin{equation}
 M_{xy}=  \pmatrix{
 {\partial U}/{\partial x}&
 {\partial U}/{\partial y}\cr
 {\partial V}/{\partial x}&
 {\partial V}/{\partial y}\cr}.
 \end{equation}
The elements of this tensor can be written via the well-known Oort constants $(A,B,C,K)_{xy},$ which describe the residual effects in our case:
 \begin{equation}
 M_{xy}=\pmatrix {K+C & A-B \cr
                  A+B & K-C \cr}.
  \label{ABCK}
  \end{equation}
According to the data presented in the last column of Table 1, we have (the matrix elements are given in km s$^{-1}$ kpc$^{-1}$)
 \begin{equation}
 M_{xy}=\pmatrix {~0.68_{(0.17)} & ~~0.0_{(0.13)} \cr
                  -0.27_{(0.17)} & -0.18_{(0.13)} \cr},\label{Mxy}
  \end{equation}
based on which we find
$A_{xy}=-0.13\pm0.11$ km s$^{-1}$ kpc$^{-1}$,
$B_{xy}=-0.13\pm0.11$ km s$^{-1}$ kpc$^{-1}$,
$C_{xy}= 0.43\pm0.11$ km s$^{-1}$ kpc$^{-1}$, and
$K_{xy}= 0.25\pm0.11$ km s$^{-1}$ kpc$^{-1}$.

Zero value of the difference $A_{xy}-B_{xy}=-\Omega_0$ and a nearly zero value of the sum of these quantities suggest that the differential Galactic rotation with the parameters (16) was taken into account very well. There is an insignificant expansion effect $K_{xy}$. $C_{xy}$ differs significantly from zero. This suggests that the residual velocity ellipse in the $xy$ plane has a deviation (vertex deviation) from the direction toward the Galactic center with $\tan 2l_{xy}=-C/A,$ then $l_{xy}=36\pm7^\circ.$ This vertex deviation can be associated both with the influence of the spiral density wave and with the influence of the warp.

{\bf The $YZ$ plane.} Consider the displacement tensor $M_{yz}$:
 \begin{equation}
 M_{yz}=  \pmatrix{
 {\partial V}/{\partial y}&
 {\partial V}/{\partial z}\cr
 {\partial W}/{\partial y}&
 {\partial W}/{\partial z}\cr}.
 \end{equation}
According to the data from the last column of Table 1, we have
 \begin{equation}
 M_{yz}=\pmatrix { -0.18_{(0.13)} & 7.36_{(1.52)} \cr
                  ~~0.25_{(0.13)} & 2.92_{(1.46)} \cr},\label{Myz}
  \end{equation}
If the rule (9) is followed strictly, then here we obtain a large negative rotation around the $x$ axis, $M_{\scriptscriptstyle32}^{\scriptscriptstyle-}=-3.55\pm0.76$~km s$^{-1}$ kpc$^{-1}$. Note that it is difficult to believe in the reality of the gradient
 $\partial V/\partial z=7.36\pm1.52$~km s$^{-1}$ kpc$^{-1}$, because it implies that the Galactic rotation velocity must increase with $z.$ In fact, exactly the reverse is true.

Bobylev (2013b) followed strictly the rule (9), when he obtained a huge value of
$\partial V/\partial z=27\pm10$~km s$^{-1}$ kpc$^{-1}$ by analyzing $\sim$200 long-period Cepheids. The linear velocity of stars with such a gradient, at $z=2$~kpc (the maximum values in our Figs. 1 and 2), can be more than 50~km s$^{-1}$. This velocity, of course, is too high. It is more likely typical for runaway stars with peculiar velocities.
As a result, Bobylev (2013b) estimated the angular velocity of rotation around the Galactic $x$ axis to be $-15\pm5$~km s$^{-1}$ kpc$^{-1}$. Cepheids within 6~kpc of the Sun, which roughly corresponds to our sample (the results of its analysis are given in the second
column of Table 1), were considered. A different Galactic rotation curve was then used to form the residual velocities of stars. The proper motions were taken from the UCAC4 (Zacharias et al. 2013) and TRC (Hog et al. 2000) catalogues. We can conclude
that a large value of the gradient $\partial V/\partial z$ is not confirmed. This gradient is most likely associated little with the actual warp rotation.

{\bf The $XZ$ plane.} Here the displacement tensor $M_{xz}$ looks as follows:
 \begin{equation}
 M_{xz}=  \pmatrix{
 {\partial U}/{\partial x}&
 {\partial U}/{\partial z}\cr
 {\partial W}/{\partial x}&
 {\partial W}/{\partial z}\cr}.
 \end{equation}
According to the data from the last column of Table 1, we have
 \begin{equation}
 M_{xz}=\pmatrix {~~0.68_{(0.17)} & -8.11_{(1.53)} \cr
                   -0.64_{(0.17)} &~~2.92_{(1.46)} \cr},\label{Mxz}
  \end{equation}
If the rule (9) is strictly followed here as well, then we obtain a large negative rotation around the $y$ axis,
$M_{\scriptscriptstyle31}^{\scriptscriptstyle-}=-3.74\pm0.77$~km s$^{-1}$ kpc$^{-1}$.

In Fig. 2 the gray color specially indicates the dependences found from the residual velocities $U$ and $V,$ which depend on whether the rotation curve and the influence of the spiral density wave are taken into account (which we disregarded). In contrast, the
dependence $\partial W/\partial x$ is represented by the red line to emphasize its importance, because the vertical velocities $W$ are not the residual ones---they are free from the influence of the Galactic rotation. This effect can be clearly seen from a comparison of the lower plots in Figs. 1g--1i and 2g--2i; the velocities $W$ remain unchanged after the correction for the Galactic rotation.

Note that the determination of the gradients $\partial U/\partial z$ and
$\partial V/\partial z$ is strongly affected by only a few stars with large $z,$ as can be seen from Figs.~2c and 2f. The actual, clearly visible dependence is seen
in Fig.~1b for the gradient $\partial U/\partial y.$

It can be seen from Fig. 2g that the vertical velocities
increase toward the Galactic anticenter (in our
case, at more negative $x$). There is a positive rotation
around the $y$ axis. Moreover, this conclusion is in
good agreement with the results of the analysis of the
vertical Cepheid velocities in Skowron et al. (2019b).

Interestingly, no three-dimensional expansion/contraction
effect $(K_{xyz})$ differing significantly from zero was detected. A noticeable two-dimensional expansion/contraction effect was not revealed in any of the
three planes either.

For the overwhelming majority of Cepheids in our sample (for 804 of the 832 stars) we know the age estimates according to the determinations by Skowron et al. (2019a). We divided the sample into two parts with an age boundary of 120~Myr. The number of relatively
young ($t<120$~Myr) and older ($t\geq120$~Myr)
Cepheids was 507 and 297, respectively. For each
of these subsamples we calculated the parameters of our model with various constraints on $R,$ but no fundamental differences, depending on the Cepheid age constraints, were found. Therefore, we do not provide the results of these calculations.

{\begin{table}[t]
 \caption[]{\small\baselineskip=1.0ex\protect
 The kinematic parameters found from the Cepheid proper motions
 }
\begin{center}
\label{t2}
\begin{tabular}{|c|r|r|r|r|}\hline
  Parameter & $R<12$~kpc & $R<14$~kpc & All Cepheids & $x<5$~kpc\\\hline

 $N_\star$        &   685 &   778 &   816 &   808 \\
 $\sigma_0$,~km s$^{-1}$ & 10.70 & 10.66 & 10.88 & 10.67 \\

 $U_\odot$,~km s$^{-1}$ & $ 6.96\pm0.51$ & $ 7.05\pm0.50$ & $ 7.20\pm0.50$ & $ 7.24\pm0.49$ \\
 $V_\odot$,~km s$^{-1}$ & $11.69\pm0.72$ & $11.06\pm0.63$ & $11.39\pm0.63$ & $11.25\pm0.63$ \\
 $W_\odot$,~km s$^{-1}$ & $ 6.12\pm0.42$ & $ 5.62\pm0.39$ & $ 5.52\pm0.39$ & $ 5.58\pm0.39$ \\
           &&&& \\
 $M_{\scriptscriptstyle32}^{\scriptscriptstyle-}$,~km s$^{-1}$ kpc$^{-1}$
           & $0.26\pm0.12$  & $ 0.17\pm0.11$ & $0.28\pm0.10$ & $0.33\pm0.10$ \\
 $M_{\scriptscriptstyle13}^{\scriptscriptstyle-}$,~km s$^{-1}$ kpc$^{-1}$
           & $0.18\pm0.22$  & $ 0.58\pm0.16$ & $0.44\pm0.14$ & $0.54\pm0.15$ \\
 $M_{\scriptscriptstyle21}^{\scriptscriptstyle-}$,~km s$^{-1}$ kpc$^{-1}$
           & $0.04\pm0.11$  & $-0.03\pm0.09$ & $0.22\pm0.09$ & $0.23\pm0.09$ \\\hline
\end{tabular}
\end{center}
\end{table}
}

\subsection*{Analysis of the Proper Motions}
It is well known (Ogorodnikov 1965) that all three rotation
components
  $M_{\scriptscriptstyle32}^{\scriptscriptstyle-},
  M_{\scriptscriptstyle13}^{\scriptscriptstyle-},
  M_{\scriptscriptstyle21}^{\scriptscriptstyle-}$ are determined
without any line-of-sight velocities, only by
analyzing the stellar proper motions. We decided to
repeat the calculations using the simplest model with
six unknowns to be determined—three components
of the peculiar solar velocity $(U,V,W)_\odot$ and three
rotation velocities
 $M_{\scriptscriptstyle32}^{\scriptscriptstyle-},
 M_{\scriptscriptstyle13}^{\scriptscriptstyle-},
 M_{\scriptscriptstyle21}^{\scriptscriptstyle-}$. In this approach
it is assumed that there are no deformations
(all components of the symmetric tensor are zero). To
be more precise, there are no deformations associated
with the warp. The conditional equations are
 \begin{equation}
  \begin{array}{lll}
  V_l= U_\odot\sin l-V_\odot\cos l
 +r[-\cos l\sin b M_{\scriptscriptstyle32}^{\scriptscriptstyle-}
    -\sin l\sin b M_{\scriptscriptstyle13}^{\scriptscriptstyle-}
    +\cos b       M_{\scriptscriptstyle21}^{\scriptscriptstyle-}],\label{eq-44}
\end{array}
 \end{equation}
 \begin{equation}
 \begin{array}{lll}
 V_b=U_\odot\cos l\sin b +V_\odot\sin l\sin b-W_\odot\cos b
 +r [\sin l M_{\scriptscriptstyle32}^{\scriptscriptstyle-}
    -\cos l M_{\scriptscriptstyle13}^{\scriptscriptstyle-}  ].
   \label{eq-55}
  \end{array}
 \end{equation}
The residual velocities are on the left-hand sides; the equations are solved by the least-squares method. The results of the solutions are presented in Table 2.

As can be seen from Fig. 2g, the vertical velocities increase toward the Galactic anticenter, consistent with the hypothesis about the warp rotation,
while in the inner Galaxy there is a large dispersion
of velocities $W.$ To exclude these Cepheids with a large dispersion of velocities $W,$ we additionally solved Eqs. (25) and (26) with a constraint on the
coordinate $x.$ The results of this solution are given in
the last column of Table~2. This solution is interesting
in that it shows the presence of two positive rotations,
each of which differs significantly from zero.

All of the solutions in Table~2 have an error per unit weight $\sim$10~km s$^{-1}$. This is less than what was obtained in the simultaneous solutions. A smaller error per unit weight is achieved not only through the nonuse of the line-of-sight velocities, but also
through a significant reduction in the number of sought-for unknowns. It should also be noted that the random errors of the line-of-sight velocities are, on average, $\sim$5~km s$^{-1}$. A typical error in the proper motion of 0.1 mas yr$^{-1}$ gives an error in the tangential velocity of 5~km s$^{-1}$ ($0.1\times 4.741\times r$) for heliocentric distances exceeding 10 kpc. Thus, in our sample the random errors of the tangential velocities are,
on average, smaller than those of the line-of-sight velocities (Fig.~1 in Bobylev et al. (2021)). It can be seen from Table~2 that there is a difference of 3--4~km s$^{-1}$ in determining the velocity $V_\odot,$ which may be associated with the peculiarity of the line-of-sight velocities. The main thing is that this effect can affect the quality of the residual velocities $V$ (when using the complete model). On the whole, we can conclude that our analysis of only the Cepheid proper motions confirms the presence of a residual rotation around the $y$ axis.

Note that the presence of a positive residual rotation around the $y$ and $x$ axes is in good agreement with the results of the analysis of a huge number of
distant stars performed by Poggio et al. (2018, 2020).

 \section*{CONCLUSIONS}
We analyzed the three-dimensional motions of a large sample of classical Cepheids. For this purpose, we used data from Mr\'oz et al. (2019). The Cepheids of this sample are located in a wide range of Galactocentric distances $R: 4-20$~kpc. The maximum elevation of these stars above the Galactic plane does not exceed 2~kpc, i.e., $|z|<2$~kpc. We applied a linear
Ogorodnikov–Milne model. The Galactic rotation that we found previously based on a nonlinear model was eliminated in advance from the observed stellar velocities.

We showed that in the Galactic xy plane there are
virtually no model components $A_{xy}$ and $B_{xy}$ differing
significantly from zero. Thus, there is no residual
rotation. Having analyzed the parameters $A_{xy}$ and
$C_{xy}$, we found that in the $xy$ plane there is a vertex
deviation $l_{xy}$ of $36\pm7^\circ.$

A slightly different situation is observed in the $zx$ and $zy$ planes. As the maximum distance $R$ of the sample increases, two gradients, $M_{13}=\partial U/\partial z$ and
$M_{23}=\partial V/\partial z$, increasingly manifest themselves. These quantities reach their maximum values at maximum radii of the neighborhood under consideration,
$M_{13}=-8.2\pm1.5$~km s$^{-1}$ kpc$^{-1}$ and
$M_{23}=7.3\pm1.5$~km s$^{-1}$ kpc$^{-1}$. We showed that the velocities $U$
and $V$ depend strongly on how the Galactic rotation curve was subtracted from them. Therefore, the values of these gradients found may not be associated with the warp rotation.

Based on a complete Ogorodnikov–Milne model, we determined the gradient
 $\partial W/\partial x=-\Omega_y$km s$^{-1}$ kpc$^{-1}$. In this case, the vertical velocity $W$ does not depend on the Galactic rotation. Therefore,
the value of $\Omega_y=+0.64\pm0.17$~km s$^{-1}$ kpc$^{-1}$
found can be interpreted as warp rotation around the $y$ axis.

The calculations based on an abridged model with only the Cepheid proper motions under the assumption of no deformations associated with the warp confirmed the presence of a residual rotation around the $y$ axis with an angular velocity
 $\Omega_y=+0.54\pm0.15$~km s$^{-1}$ kpc$^{-1}$. They also showed the presence
of a slight positive rotation around the x axis with an
angular velocity $\Omega_x=+0.33\pm0.10$~km s$^{-1}$ kpc$^{-1}$.

\bigskip{\bf ACKNOWLEDGMENTS}

We are grateful to the referee for the useful remarks that contributed to an improvement of the paper.

\bigskip \bigskip\medskip{\bf REFERENCES}{\small

1. I. Ablimit, G. Zhao, C. Flynn, and S. A. Bird, Astrophys. J. 895, L12 (2020).

2. R. I. Anderson, H. Saio, S. Ekstr\"om, C. Georgy, and G. Meynet,
Astron. Astrophys. 591, A8 (2016).

3. J. Bailin, Astrophys. J. Lett. 583, L79 (2003).

4. E. Battaner, E. Florido, andM. L. Sanchez-Saavedra, Astron. Astrophys. 236, 1 (1990).

5. K. Bekki, Mon. Not. R. Astron. Soc. 422, 1957
(2012).

6. L. N. Berdnikov, Astron. Lett. 13, 45 (1987).

7. V. V. Bobylev, Astron. Lett. 36, 634 (2010).

8. V. V. Bobylev, Astron. Lett. 39, 753 (2013a).

9. V. V. Bobylev, Astron. Lett. 39, 819 (2013b).

10. V. V. Bobylev, A. T. Bajkova, A. S. Rastorguev, and M. V. Zabolotskikh, Mon. Not. R. Astron. Soc. 502, 4377 (2021).

11. A. G. A. Brown, A. Vallenari, T. Prusti, J. H. J. de Bruijne, C. Babusiaux, Bailer-Jones, M. Biermann, D. W. Evans, et al. (Gaia Collab.),
Astron. Astrophys. 616, 1 (2018).

12. J. C. Cersosimo, S. Mader, N. S. Figueroa, et al., Astrophys. J. 699, 469 (2009).

13. X. Cheng, B. Anguiano, S. R. Majewski, C. Hayes, P. Arras, C. Chiappini, S. Hasselquist, Q. de Andrade, et al., Astrophys. J. 905, 49 (2020).

14. Z. Chrob\'akov\'a and M. L\'opez-Corredoira, Astrophys.
J. 912, 130 (2021).

15. S. V. M. Clube, Mon. Not. R. Astron. Soc. 159, 289 (1972).

16. R. Drimmel, R. L. Smart, and M. G. Lattanzi, Astron. Astrophys. 354, 67 (2000).

17. R. Drimmel and D. N. Spergel, Astrophys. J. 556, 181 (2001).

18. J. D. Fernie, Astron. J. 73, 995 (1968).

19. The HIPPARCOS and Tycho Catalogues, ESA SP--1200 (1997).

20. E. Hog, C. Fabricius, V. V. Makarov, et al., Astron. Astrophys. 355, L27 (2000).

21. P. M. W. Kalberla and L. Dedes, Astron. Astrophys. 487, 951 (2008).

22. M. L\'opez-Corredoira, J. Betancort-Rijo, and J. Beckman,
Astron. Astrophys. 386, 169 (2002a).

23. M. L\'opez-Corredoira, A. Cabrera-Lavers, F. Garz\'on, and P. L. Hammersley,
Astron. Astrophys. 394, 883 (2002b).

24. M. L\'opez-Corredoira, H. Abedi, F. Garz\'on, and
F. Figueras, Astron. Astrophys. 572, 101 (2014).

25. M. Miyamoto and Z. Zhu, Astron. J. 115, 1483 (1998).

26. Y. Momany, S. Zaggia, G. Gilmore, et al., Astron. Astrophys. 451, 515 (2006).

27. P. Mr\'oz, A. Udalski, D. M. Skowron, et al., Astrophys. J. 870, L10 (2019).

28. K. F. Ogorodnikov, Dynamics of Stellar Systems, Ed. by A. Beer (Pergamon, Oxford, 1965; Fizmatgiz, Moscow, 1965).

29. C. A. Olano, Astron. Astrophys. 423, 895 (2004).

30. E. Poggio, R. Drimmel, M. G. Lattanzi, R. L. Smart, A. Spagna, R. Andrae, C. A. L. Bailer-Jones, M. Fouesneau, et al., Mon. Not. R. Astron. Soc. 481, L21 (2018).

31. E. Poggio, R. Drimmel, R. Andrae, C. A. L. Bailer-Jones, M. Fouesneau, M. G. Lattanzi, R. L. Smart, and A. Spagna, Nat. Astron. 4, 590 (2020).

32. S. R\"oser, M. Demleitner, and E. Schilbach, Astron. J. 139, 2440 (2010).

33. D. Russeil, Astron. Astrophys. 397, 133 (2003).

34. R. Sch\"onrich, J. J. Binney, andW. Dehnen, Mon. Not.
R. Astron. Soc. 403, 1829 (2010).

35. D. M. Skowron, J. Skowron, P. Mr\'oz, et al., Science
(Washington, DC, U.S.) 365, 478 (2019a).

36. D. M. Skowron, J. Skowron, P. Mr\'oz, et al., Acta Astron. 69, 305 (2019b).

37. M. F. Skrutskie, R. M. Cutri, R. Stiening, M. D. Weinberg, S. Schneider, J. M. Carpenter, C. Beichman, R. Capps, et al., Astron. J. 131, 1163 (2006).

38. L. Sparke and S. Casertano, Mon. Not. R. Astron. Soc. 234, 873 (1988).

39. T. Tsuchiya, New Astron. 7, 293 (2002).

40. A. Udalski, M. K. Szyma\'nski, and G. Szyma\'nski, Acta Astron. 65, 1 (2015).

41. S. Wang, X. Chen, R. de Grijs, et al., Astrophys. J. 852, 78 (2018).

42. G. Westerhout, Bull. Astron. Inst. Netherlands 13, 201 (1957).

43. I. Yusifov, astro-ph/0405517 (2004).

44. N. Zacharias, C. Finch, T. Girard, et al., Astron. J. 145, 44 (2013).

}
\end{document}